\documentclass[times]{qjrms4}

\usepackage[colorlinks,bookmarksopen,bookmarksnumbered,citecolor=red,urlcolor=red]{hyperref}
\usepackage[normalem]{ulem}
\usepackage{graphicx}
\usepackage{natbib}
\usepackage{color}
\usepackage{mathtools}
\usepackage{latexsym}
\usepackage{subfig}
\usepackage{natbib}
\usepackage{upgreek}
\usepackage{mathrsfs}
\usepackage{float}
\usepackage{caption}
\usepackage{soul}
\usepackage{textcomp}
\usepackage{stmaryrd}
\usepackage{amsmath}
\usepackage{lipsum}                     % Dummytext
\usepackage{xargs}                      % Use more than one optional parameter in a new commands
\usepackage[pdftex,dvipsnames]{xcolor}  % Coloured text etc.
\usepackage[colorinlistoftodos,prependcaption,textsize=small]{todonotes}

 \def\der#1#2{{\partial #1\over \partial #2}}
 \def\deriv#1#2{{d #1\over d #2}}

\def\be{\begin{equation}}
\def\ee{\end{equation}}
\def\bea{\begin{eqnarray}}
\def\eea{\end{eqnarray}}
\def\bse{\begin{subequations}}
\def\ese{\end{subequations}}
\def\bsea{\begin{subeqnarray}}
\def\esea{\end{subeqnarray}}
\def\({\left (}
\def\){\right )}
\def\[{\left [}
\def\]{\right ]}
\def\{{\left\{}
\def\}{\right\}}
\def\<{\left <}
\def\>{\right >}

\usepackage{cleveref}

\crefformat{section}{section #2#1#3} % see manual of cleveref, section 8.2.1
\crefformat{subsection}{section #2#1#3}
\crefformat{subsubsection}{section #2#1#3}

\newcommand\BibTeX{{\rmfamily B\kern-.05em \textsc{i\kern-.025em b}\kern-.08em
T\kern-.1667em\lower.7ex\hbox{E}\kern-.125emX}}

\usepackage{moreverb}

\def\volumeyear{2013}

\begin{document}
\runningheads{Heifetz,~Maor~and~Guha}{Boussinesq and non-Boussinesq baroclinic torques of surface gravity wave propagation}
%\runningheads{A.~N.~Other}{A demonstration of the \emph{Q.~J.~R.
%Meteorol. Soc.} class file}

\title{On the opposing roles of the Boussinesq and non-Boussinesq baroclinic torques in surface gravity wave propagation}

\author{%%%% Author details
Eyal Heifetz$^{1}$, Ron Maor$^1$  and Anirban Guha$^{2,3}$\corrauth}

%%%%%%%%% Insert author address here
\address{$^{1}$Department of Geophysics, Porter school of the Environment and Earth Sciences, Tel Aviv University,
Tel Aviv 69978, Israel.\\
$^{2}$ School of Science and Engineering, University of Dundee, Nethergate, Dundee DD1 4HN, U.K.\\
$^{3}$Institute of Coastal Research, Helmholtz-Zentrum Geesthacht, Geesthacht, 21502 Germany.
}
\corraddr{Email: anirbanguha.ubc@gmail.com}

\begin{abstract}

%should be rewritten 

%The standard derivation of the surface gravity waves dispersion relation, at the ocean -- atmosphere interface, invokes the potential flow kinematic and dynamic matching conditions. This derivation however seems to obscure the mechanistic understanding of the wave propagation. 

Here we suggest an alternative understanding of the surface gravity wave propagation mechanism based on the baroclinic torque, which operates to translate the interfacial vorticity anomalies at the air-water interface. We demonstrate how the non-Boussinesq term of the baroclinic torque acts against the Boussinesq one to hinder wave propagation. By standard vorticity inversion and mirror imaging, we then show how the existence of the bottom boundary affects the two types of torque. Since the opposing non-Boussinesq torque results solely from the  mirror image, it vanishes in the deep water limit and its magnitude is half of the Boussinesq torque in the shallow water limit. This reveals that Boussinesq approximation is valid in the deep water limit, even though the density contrast between air and water is large. The mechanistic roles, played by the Boussinesq and non-Boussinesq parts of the baroclinic torque, remain obscured in the standard derivation where the time-dependent Bernoulli equation is implemented instead of the interfacial vorticity equation.
Finally, we note on passing that the Virial theorem for surface gravity waves can be obtained solely from considerations of the dynamics at the air-water interface.  

%on the vortex sheet dynamics at the air-water interface. We examine how the two terms of the baroclinic torque operate to translate the interfacial vorticity anomalies and how the non-Boussinesq term of the baroclinic torque acts against the Boussinesq one to hinder the wave propagation. We further investigate, via vorticity inversion, how the existence of the bottom boundary affects the two types of torque by replacing this boundary by an anti-phased vortex sheet mirror image. Since the opposing non-Boussinesq torque results solely from the  mirror image, it vanishes in the deep water limit and its magnitude is half of the Boussinesq torque in the shallow water limit. Finally, we demonstrate how the Virial theorem for surface gravity waves can be obtained only from considerations of the dynamics at the wave interface. 

\end{abstract}

%\pacs{47.20.Ma,47.15.St,47.20.Ft}% PACS, the Physics and Astronomy
\keywords{surface gravity waves, baroclinic torque, non-Boussinesq flows}%Use showkeys class option if keyword

%\pacs{47.52.+j, 47.15.St, 05.45.Xt, 47.20.Ky}

\maketitle

\section{Introduction}

The fundamental theory of linear surface gravity waves was founded in the 18th and the early 19th centuries by pioneers like Laplace, Lagrange, Poisson, and Cauchy \citep{craik2004origins}. The standard derivation of the dispersion relation, based on the potential flow theory, is a well known procedure considered in almost all standard textbooks in fluid dynamics (e.g. see \citet{kundu2016fluid}).

%In this derivation, however, the implementation of the kinematic and dynamic conditions 
%(the latter via the time-dependent Bernoulli equation) seems to obscure the physical mechanism of the wave propagation and the reason for the peculiar dispersion relation $\omega(k) = \pm \sqrt{gk\tanh{(kH)}}$ (where $\omega$ denotes frequency, $g$ is gravity, $k$ is the wavenumber and $H$ is the water depth). 

Recently, there has been a growing body of literature
\citep{holmboe1962behavior,sakai1989rossby,baines1994mechanism,caulfield1994multiple,harnik2008,rabinovich2011,carp2012,guha2014wave,heifetz2015stratified} 
%\textcolor{orange}{my relevant references for stratified shear flows are not heifetz1999counter,heif2005 but Harnik et al, Rabinovich et al and Heifetz and Mak}
dealing with stratified shear flow instability  that treats the dynamics of density discontinuity surfaces as \emph{interfacial vorticity waves}\footnote{We define any linear interfacial wave that propagates due to vorticity anomalies across the interface as a vorticity wave. Hence Rossby waves, gravity waves, capillary waves, Alfv\'en waves are all vorticity waves by this definition.}. This approach provides a mechanistic rationalization for Taylor-Caulfield and Holmboe instabilities, and also paves the path for efficient vortex-method-based computation schemes to simulate the nonlinear evolution, including wave-breaking \citep{bhardwaj2018nonlinear}. However, while exploring these relatively complex problems, the analysis of surface gravity waves - probably the simplest setup of a density discontinuity - in terms of interfacial vorticity waves has been ``left behind''. Hence, here we suggest an alternative derivation of surface gravity waves, based on the above-mentioned approach. To avoid confusion, we emphasize here that this work uses standard techniques of vorticity inversion and boundary mirror images in order to focus on the wave propagation mechanism rather than on the modelling of surface waves as vortex sheets, a well-known technique pioneered by \cite{baker1982generalized}.

%The intrinsic phase speed $c = \pm \sqrt{(g/k)\tanh{(kH)}}$ (where $g$ is gravity, $k$ is the wavenumber and $H$ is the water depth) obtained from this linear analysis is considered to be the fundamental speed of surface gravity waves even when nonlinear wave interaction and shoaling are modeled (e.g. Toledo 2013, Guha 2018). Despite of the central role of this dispersion relation, it is quite difficult to obtain a physical mechanistic understanding to the particular dependence of the phase speed in the wavenumber and the role played by the solid bottom boundary, especially in light of the widely used approximations of deep and shallow-water. The implementation of the kinematic and dynamic conditions (the latter via the time dependent Bernoulli equation) in the standard procedure, rather than an explicit use of the equation of motions seem to obscure the physical mechanism of the wave propagation. It is our goal here is to provide an alternative derivation, based solely on the vortical dynamics of the air-water interface, in hope that this approach will provide additional understanding to this basic phenomena. 

This article is organized as follows. In section \ref{sec2}, we derive and analyze the interfacial vorticity equations and then in section \ref{sec3}, we implement them to obtain the dispersion relation. Special care is given to the deep and the shallow-water limits. In section \ref{sec4} we derive surface waves' energy from the interfacial fields. Finally, we close by discussing the results in section \ref{sec5}.

\section{Interfacial vorticity dynamics}
\label{sec2}
Assuming an incompressible, inviscid flow the governing continuity and momentum equations read 
\renewcommand{\theequation}{\arabic{section}.\arabic{equation}a,b}
\begin{equation}
\nabla \cdot \mathbf{u}=0,\,\,\, \frac{D\rho}{Dt}=0,
\label{eq:gov_NSC_1}
\end{equation} 
\addtocounter{equation}{-1}
\renewcommand{\theequation}{\arabic{section}.\arabic{equation}c}
\begin{equation} 
\frac{D\mathbf{u}}{Dt}=-g\hat{\mathbf{k}}-\frac{1}{\rho}\nabla p,
\label{eq:gov_NSC_2}
\end{equation} 
where $D/Dt \equiv \partial/\partial t + \mathbf{u}\cdot \nabla$ is the material derivative,  $\mathbf{u}$, $\rho$, $p$, $g$ and  $\mathbf{\hat{k}}$ respectively denote velocity, density, pressure, gravity and the vertical unit vector. The vorticity equation is  obtained by taking the curl of Equation\ \ref{eq:gov_NSC_2}: 
\renewcommand{\theequation}{\arabic{section}.\arabic{equation}}
\begin{equation}
\frac{D\boldsymbol{\omega}}{Dt}=\underbrace{ (\boldsymbol{\omega} \cdot \nabla)\mathbf{u}}_\text{\clap{Stretching}}  +\underbrace{\frac{1}{\rho^2}\nabla\rho\times\nabla p}_\text{\clap{Baroclinic}},
\label{eq:full-vort}
 \end{equation}
where $\boldsymbol{\omega}\equiv \nabla \times \mathbf{u}$ is the vorticity. The first term on the RHS is the vortex stretching term that would be absent in 2D flows. The second term signifies the baroclinic torque when isopycnals and isobars are crossing. In general, this term can be divided into two parts --  \emph{Boussinesq} and  \emph{non-Boussinesq}.

% We consider a 2D flow $\mathbf{u}=(u,\,w)$ in the $x-z$ plane, and assume a base state that varies only along $z$: $\bar{u}=\bar{u}(z)$, $\bar{w}=0$, $\bar{q}(z)=d\bar{u}/dz$, $\bar{p}=\bar{p}(z)$ and $\bar{\rho}=\bar{\rho}(z)$. The base state follows hydrostatic pressure balance $d\bar{p}/dz=-\bar{\rho}g$. We add perturbations to the base flow and the governing equations are linearized. This produces the perturbation vorticity evolution equation

Consider now the 2D $(x,z)$ setup of hydrostatic background
\be
{d\overline p \over dz} = -\overline{\rho}g\,,
\label{eq:hydrostatic}
\ee
where $\overline{p}$ and $\overline{\rho}$ respectively denote the mean pressure and density. Linearisation with respect to Equations \ref{eq:hydrostatic}, \ref{eq:gov_NSC_1} and \ref{eq:gov_NSC_2}  yield
\renewcommand{\theequation}{\arabic{section}.\arabic{equation}a}
\be
\label{eq:3d}
\der{u'}{x}+\der{w'}{z}=0,
\ee
\addtocounter{equation}{-1}
\renewcommand{\theequation}{\arabic{section}.\arabic{equation}b}
\be
\label{eq:3c}
\der{\rho'}{t} = -w'\frac{d{\overline\rho}}{dz},
\ee
\renewcommand{\theequation}{\arabic{section}.\arabic{equation}a}
\be
\label{eq:3a}
\der{u'}{t}=-\frac{1}{\overline \rho}\der{p'}{x}\, ,
\ee
\addtocounter{equation}{-1}
\renewcommand{\theequation}{\arabic{section}.\arabic{equation}b}
\be
\label{eq:3b}
\der{w'}{t}=-\frac{1}{\overline\rho}\der{p'}{z} -{\rho' \over \overline\rho}g,
\ee
where the primes denote perturbation quantities. { \emph{ Hereafter, the primes will be dropped so that any variable without an overbar will denote a perturbation quantity}. }

\begin{figure}
\centering
\includegraphics[width=1.0\linewidth]{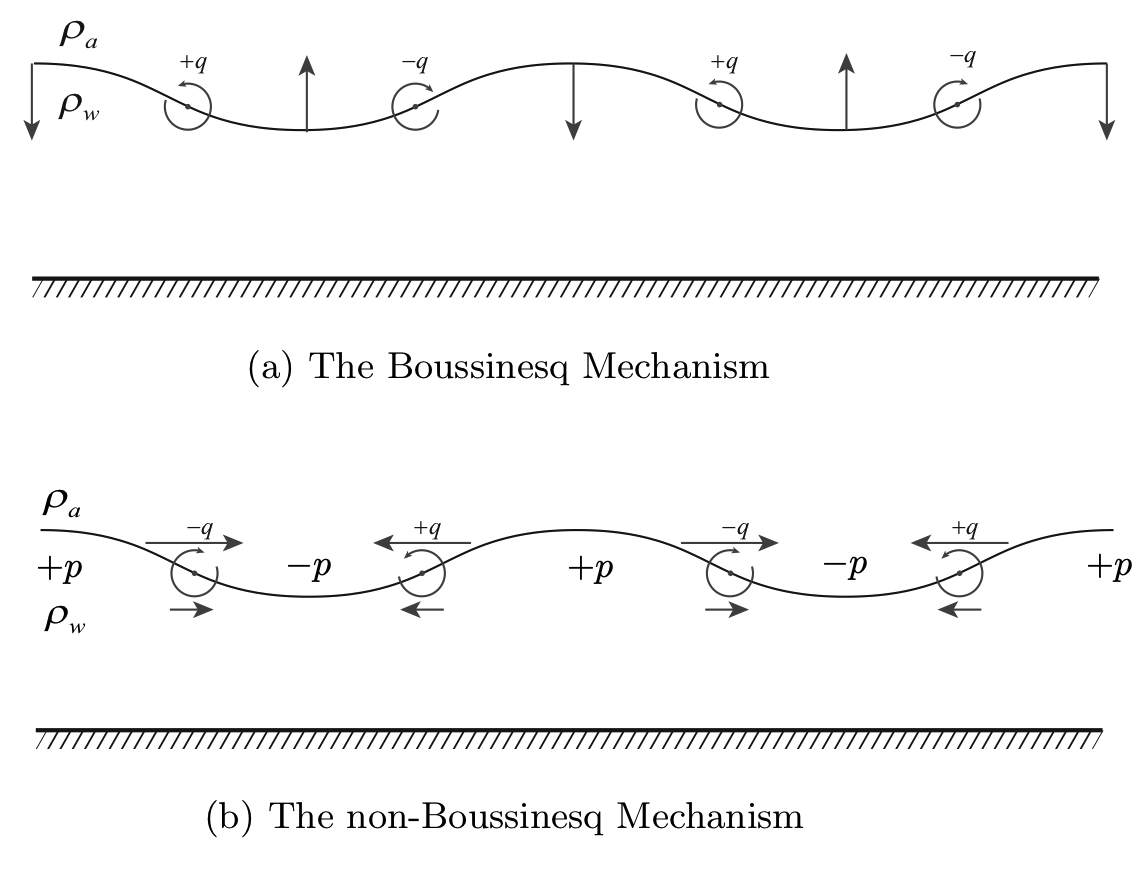}
\caption{Schematic illustration of the air-water interface (wavy line); density of air and water are respectively $\rho_a$ and $\rho_w$.  (a) The vertical arrows denote  vertical velocity generated by the Boussinesq term  ($T_1$), contributing to the  $\partial w/\partial x$ in $q$. (b) The non-Boussinesq term ($T_2$) is a scaled horizontal perturbation pressure gradient force generating vertical shear of the horizontal velocity, $ - \partial u/\partial z$, and thus contributes to the generation of  $q$. In both (a) and (b), the generated $q$ is shown by circular arrows. 
% of $q$ at the crests of $\eta$ is marked.  The shaded regions on the
% top plot (at time $t=t_0$) denote regions of positive vorticity
% generation and negative displacement generation. This vorticity and
% displacement generation patter shifts the wave westward relative to
% the mean flow (denoted by the thick horizontal arrow). The wave
% position at a quarter period later is shown below (marked
% $t=t_0+\Delta t$).  For this case there is only the negatively
% correlated $\zm$ mode (see text for details), which is a westward
% propagating Rossby wave (for $\qz>0$).
}
\label{fig:1}
\end{figure}

Defining the 2D vorticity\footnote{$q$ points in the negative $y$ direction in order to follow the convention in which counterclockwise (clockwise) rotation is associated with positive (negative) values of vorticity.} perturbation as
\renewcommand{\theequation}{\arabic{section}.\arabic{equation}}

\be
q = \der{w}{x} - \der{u}{z},
\ee
the 2D version of Equation\,\ref{eq:full-vort} under hydrostatic balance yields
\begin{equation}\label{eq:vort1}
\der{q}{t}= \underbrace{-\frac{g}{\bar{\rho}}\frac{\partial  {\rho}}{\partial x}}_\text{\clap{$T_1\equiv$Boussinesq baroclinic}}\,\,\,\,\,\,\,\,\,\,+\,\,\,\,\,\,\,\,\,\,\, \underbrace{\left[-\frac{1}{\bar{\rho}^{2}}\Big(\frac{d \bar{\rho}}{d z}\frac{\partial  {p}}{\partial x} \Big)\right].}_\text{\clap{$T_2\equiv$Non-Boussinesq baroclinic}}
\end{equation}
The two terms in the RHS of Equation \ref{eq:vort1} emanate from the baroclinic term appearing in the RHS of Equation
\ref{eq:full-vort}. { Both $T_1$ and $T_2$ point in the negative $y$ direction, in agreement with the definition of $q$.} The vorticity can be generated from two baroclinic sources:
\\(i) $T_1 \equiv -(g/\bar{\rho}){\partial \rho}/{\partial x}$: denotes  the ``Boussinesq baroclinic torque''. 
It results from the buoyancy restoring force acting to flatten a displaced density interface to its initial unperturbed horizontal position. Therefore, it generates the horizontal shear of the vertical velocity, $\partial w/\partial x$, in $q$; see Figure \ref{fig:1}a.
%For example, at the warm-water--cold-water interface (thermocline),  this term  would be responsible for the propagation of interfacial gravity waves. In fact, it is the only baroclinic generation term present when the Boussinesq approximation is invoked.   
%Under   Boussinesq approximation ($\bar{\rho} \approx \mathrm{const.}$), this term signifies the Boussinesq baroclinic generation and underpins the key mechanism behind internal gravity wave propagation.
\\(ii) $T_2\equiv -(1/\bar{\rho}^{2})(d \bar{\rho}/{d z})({\partial p}/{\partial x})$: denotes  the ``non-Boussinesq baroclinic generation of vorticity'', arises out of the density variations in the inertial terms, and is completely independent of the gravitational effects. This term vanishes if Boussinesq approximation (which, in a simplified sense, implies $\bar{\rho} \approx \mathrm{const.}$ in Equation\ \ref{eq:vort1}) is invoked. Alternatively, if we assumed  Boussinesq approximation from the onset, then Equations \ref{eq:3a}--\ref{eq:3b} would have resulted in Equation\ \ref{eq:vort1} with $T_2 \equiv 0$.  
$T_2$ results from the decrease with height of the mean density ${\overline\rho}$. Consequently, a horizontal perturbation pressure gradient accelerates the upper lighter material faster than the heavier material below it. 
This accounts to the vertical shear of the horizontal velocity, $ - \partial u/\partial z$, in $q$; see Figure \ref{fig:1}b (a more detailed analysis can be found in \citet{heifetz2015stratified}).

Note that under situations when the displacement and the pressure perturbations correlate, $T_1$ and $T_2$ act against each other in generating vorticity. This will be shown to be the case for the air-water interface -- for surface gravity waves over water of finite depth, a lift up of the interface is associated with a positive density anomaly and thus with a positive pressure perturbation.

% {
% We denote the first term in the RHS as the Boussinesq term and the second as the non-Boussinesq one, since the latter is not appeared in the Boussinesq equations where ${\overline\rho}(z)$ is replaced by a constant reference density value in the denominator of the pressure gradient force in (3a,b).}

The linearized kinematic condition at an interface (e.g. air-water free surface) is given by
\renewcommand{\theequation}{\arabic{section}.\arabic{equation}}
\be
\label{eq:4}
\der{\eta}{t} = w\, ,
\ee
where $\eta$  denotes the free surface displacement. One can combine Equations \ref{eq:3c} and \ref{eq:4} to obtain
\be
\label{eq:dens}
\rho = -\eta \frac{d\overline\rho}{dz}.
\ee
Equations \ref{eq:dens} and \ref{eq:vort1} can be combined together, yielding
%(3c,d) suggest that if all density perturbations result from vertical advection of the mean density then $\rho = -\eta \der{\overline\rho}{z}$. This allows us to write (5) as:
\be
\der{q}{t} = -N^2\der{}{x}\(\eta- {p\over {\overline\rho}g} \)\, ,
\label{eq:2.09}
\ee      
where $N(z) \equiv \sqrt{-({g}/{\overline\rho}) {d\overline\rho}/{dz}}$ is the Brunt-V\"ais\"al\"a frequency.
The RHS of Equation \ref{eq:2.09} indicates that close to hydrostaticity of the perturbations, $T_1$ and $T_2$ are of the same order of magnitude. 
%where $N^2 = -{g \over {\overline\rho}}\der{\overline\rho}{z}\,$ is the square of the Brunt-Vaisala frequency.
%In Figure \ref{fig:1} we illustrate how the two terms in the RHS generates vorticity for stable stratification (positive $N^2$). The term $T_1$ results from the buoyancy restoring force acting to flatten a displaced density interface to its initial unperturbed horizontal position. Therefore, it generates the horizontal shear of the vertical velocity, $\partial w/\partial x$, in $q$; see Figure \ref{fig:1}a. The $T_2$ term results from the decrease with height of the mean density ${\overline\rho}$. Consequently, a horizontal perturbation pressure gradient accelerates the upper light material faster than the heavy material below it. 
%This accounts to the vertical shear of the horizontal velocity, $ - \partial u/\partial z$, in $q$; see Figure \ref{fig:1}b (a more detailed analysis can be found in \citet{heifetz2015stratified}). Note from Equation \ref{eq:2.09} that under situations when the displacement and the pressure perturbations correlate, $T_1$ and $T_2$ can act against each other in generating vorticity. This will be shown to be the case for the air-water interface -- for surface gravity waves over water of finite depth, a lift up of the interface is associated with a positive density anomaly and thus with a positive pressure perturbation. Furthermore, 
Using Equation \ref{eq:3a} we can rewrite Equation
\ref{eq:2.09} as:    
\be
\label{eq:2.10}
\der{}{t}\(q +{N^2 \over g}u\) = -N^2\der{\eta}{x}\, , 
\ee
where we note that the term $T_2$ is now the second term of the LHS. %{We emphasize here that, while the standard derivation of the dynamic condition (i.e., the time-dependent Bernoulli's equation) fully relies on  potential flow theory, c.f. \cite{kundu2016fluid},  our dynamic condition Equation \ref{eq:2.10} \emph{does not} require any such assumption. Hence Equation \ref{eq:2.10} should not be confused with the time-dependent Bernoulli's equation.
%We also note in passing that, although the time-dependent Bernoulli's equation in water wave theory is almost always applied \emph{at} the interface, e.g. see  \cite{kundu2016fluid}, the technically correct application of it should be just below the interface since the latter itself is a vortex sheet. In summary, Equation \ref{eq:2.10} is valid \emph{at} the interface while the time-dependent Bernoulli's equation is valid \emph{just below} the interface.
%}

\subsection{Surface gravity waves}
Consider now the standard setup of the air-water interface,  at rest at $z=0$. The water depth is $H$, while the air above is unbounded. The mean density can be written as
\be
{\overline\rho}(z) = 
\begin{cases}
\rho_a \rightarrow 0\, , &  0 < z <  \infty, \\
\rho_w\, ,& -H <z <0,
\end{cases}\, 
\ee
yielding $N^2 = 2g\delta(z)$, where $\delta$ denotes the Dirac delta function (if $\rho_a$ is not neglected, $g$ would be replaced by the reduced gravity). Equation \ref{eq:2.10} therefore implies  that the vorticity perturbation will be generated only at the air-water interface, giving rise to an undulating vortex sheet. We wish to describe the gravity wave dynamics solely from the vorticity dynamics of this interface. For modal solution of the form
\be
\label{eq:2.12}
q = {\hat q}_0\delta(z)e^{ik(x-ct)},   
\ee
where the subscript `$0$' indicates evaluation at $z=0$, ${\hat q}_0$ denotes the vorticity amplitude and $c$ is the phase speed. The vorticity can be inverted to obtain the streamfunction $\psi = \nabla^{-2} q$ to express the velocity field $$(u,w) = \left(-\der{}{z},\der{}{x}\right)\psi.$$ 
Using the Green function approach we can write:
\begin{align}
    \psi = \nabla^{-2} q & = \int_{-H}^{\infty} q\, G(z,z',k)dz' \nonumber\\
    &=  {\hat q}_0  G(z,0,k)e^{ik(x-ct)} =  {\hat \psi}(z)e^{ik(x-ct)},
\end{align}
% \psi = \nabla^{-2} q = \int_{-H}^{\infty} q\, G(z,z',k)dz' =  {\hat q}_0  G(z,0,k)e^{ik(x-ct)} =  {\hat \psi}(z)e^{ik(x-ct)}\, ,
% \ee
where the Green function $G$ { satisfies}  Helmholtz-Poisson's equation:
\be
\label{eq:Green_delta}
\(-k^2  +\der{^2}{z^2}\) G(z,0,k) =  \delta(z).
\ee
The structure of the Green function results from the assumption of potential flow dynamics above and below the air-water discontinuity, where the jump in $u$ across the interface yields the vorticity delta function  of Equation \ref{eq:2.12}.
Furthermore, we apply the evanescent condition $w(z\rightarrow \infty) \rightarrow 0$, and the solid impenetrable boundary condition, $w(z=-H) = 0$, implying $G(z= \infty, 0,k) = G(z=-H, 0,k)= 0$. 

%Equations \ref{eq:2.12}-\ref{eq:Green_delta} are equivalent to the conventional dynamic condition obtained from the matching condition of the time-dependent Bernoulli's equation. 

\begin{figure}
\centering
\includegraphics[width=1.0\linewidth]{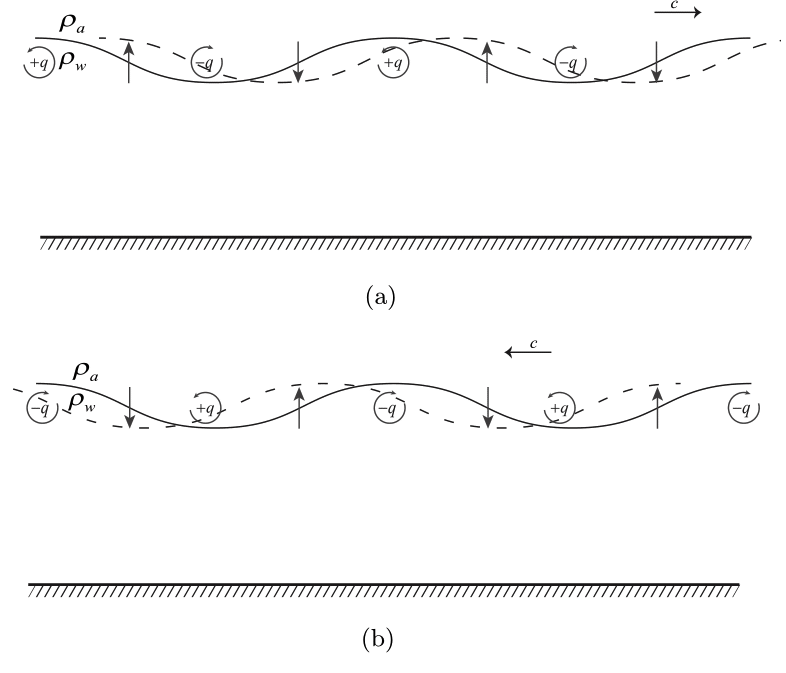}
\caption{Schematic illustration of the propagation of  surface gravity waves. (a) When the interface vorticity (indicated by circular arrows) and
displacement (indicated by the solid undulating line) are in phase, the vertical
velocity (indicated by the vertical arrows) induced by the  perturbation vorticity is located a quarter of wavelength to the right of the displacement, thus
translating the displacement (the dashed undulating line) in the positive $x$ direction (Equation \ref{eq:2.15a}). (b) Same as in (a), except that now the wave propagates in
the negative $x$ direction as the vorticity and the displacement anomalies are in
anti-phase.
}
\label{fig:2}
\end{figure}

The dynamics finally boils down to finding the modal solution that  simultaneously satisfies Equation \ref{eq:4} and Equation \ref{eq:2.10} at the interface. { Writing 
$\eta = {\hat \eta}_0 e^{ik(x-ct)}$ and recall that 
$N^2 = 2g\delta(z)$, we obtain}
% \renewcommand{\theequation}{\arabic{section}.\arabic{equation}a,b}
% \be
% \label{eq:2.15}
% c = {i\over k}{{\hat w}_0  \over {\hat \eta}_0}\, ; \hspace{0.5cm}
% c= {2g{\hat \eta}_0 \over {\hat q}_0 + 2{\hat u}_0}\, ,
% \ee
\renewcommand{\theequation}{\arabic{section}.\arabic{equation}a}
\be
\label{eq:2.15a}
c = {i\over k}{{\hat w}_0  \over {\hat \eta}_0},
\ee
\addtocounter{equation}{-1}
\renewcommand{\theequation}{\arabic{section}.\arabic{equation}b}
\be
\label{eq:2.15b}
c= {2g{\hat \eta}_0 \over {\hat q}_0 + 2{\hat u}_0},
\ee
\renewcommand{\theequation}{\arabic{section}.\arabic{equation}}

\noindent where both $({\hat u}_0, {\hat w}_0)$ are expressed in terms of ${\hat q}_0$. Equation \ref{eq:2.15a}, which results from the kinematic condition \ref{eq:4}, reveals that for  waves propagating in the positive (negative) $x$ direction, the vertical velocity is located a quarter of wavelength to the right (left) of the displacement anomaly (see Figure \ref{fig:2}). Equation \ref{eq:2.15b}, which results from the interfacial vorticity equation \ref{eq:2.10}, indicates that 
$( {q}_0 + 2{u}_0)$ and $\eta_0$ should be in phase (anti-phase) for positive (negative) wave propagation. While the former is a general property of any transverse wave, the latter will be clarified in the next section. { Equation \ref{eq:2.15b} can be obtained as well from the conventional dynamic condition. Writing the time dependent Bernoulli equation for the potential flow just below the interface: 
\be
\label{Bernoulli}
-\der{\phi}{t} = {{\bf u}^2 \over 2} + {p\over \rho} +gz,
\ee
where the velocity potential $\phi$ satisfies ${\bf u} = \nabla \phi$,  we obtain after linearisation that just below the wavy interface 
$\partial \phi/\partial t = -g\eta$, so that 
$c= {g{\hat \eta}_0 / ik{\hat \phi}_0} $, where  $ik{\hat \phi}_0 =  {\hat q}_0/2 + {\hat u}_0$. The first term in the RHS is the Boussinesq contribution to the velocity potential, whereas the second is the non-Boussinesq one. Nonetheless, the dynamic condition by itself does not allow us to  { straightforwardly separate}  the Boussinesq and the Non-Boussinesq dynamics.  }

We also note in passing an important aspect of transverse waves by simply analyzing Equation \ref{eq:2.15a}. The maximum (or minimum) phase speed occurs when $\partial c/\partial k =0$, which gives for a reference  wave amplitude displacement ${\hat \eta}_0$:
\be
\frac{\partial{\hat w}_0}{\partial k}=\frac{{\hat w}_0}{k} \implies {\hat w}_0 =  \alpha k,
\ee
where $\alpha$ is a constant. Substituting this relation in Equation \ref{eq:2.15a} yields
\be
\label{eq:cmax}
c_{max} = \bigg|{\alpha  \over {\hat \eta}_0}\bigg|.
\ee
The above equation shows that the maximum phase speed is \emph{independent} of the wavenumber $k$, implying that in the fast phase speed limit, transverse waves are non-dispersive. {It is important to note that the above relation is  \emph{independent} of the dynamic condition and hence applicable to any transverse wave.}
This simple derivation provides a mathematical justification of  the well-known fact that in the shallow water limit, the phase speed is maximum and the  waves are non-dispersive.

\section{Dispersion relation analysis}
\label{sec3}
\subsection{deep-water limit}
\label{subsec3_1}
\setcounter{equation}{0}
%\addtocounter{equation}{-15}

\begin{figure}
\centering
\includegraphics[width=1.0\linewidth]{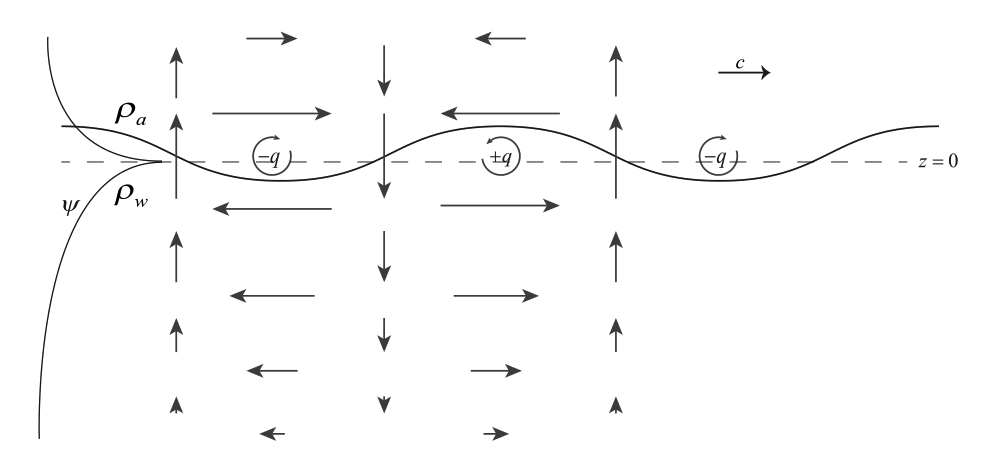}
\caption{The velocity field of the rightward propagating surface gravity wave in the
deep-water limit. The vertical structure of the deep-water streamfunction  (Equation \ref{eq:3.1}) is
illustrated in the left. Due to its evanescent structure away from the interface, the flow is everywhere non-divergent. It is also irrotational everywhere except at
the interface, where the jump in the horizontal velocity across the interface generates
a  delta function in vorticity. For leftward propagating waves in the deep
water limit, the velocity field is the same but the displacement is in anti-phase
with the vorticity anomaly.
}
\label{fig:3}
\end{figure}

Consider first the deep-water (hereafter DW) limit, that is, $kH\rightarrow \infty$. In this case, the Green function $G_{DW} = -e^{-k|z|}/2k$, which yields 
\renewcommand{\theequation}{\arabic{section}.\arabic{equation}}
\be
\label{eq:3.1}
%\[\psi = -{{\hat q}_0 \over 2k}e^{-k|z|}e^{ik(x-ct)}\]_{DW}\, .
\psi = -{{\hat q}_0 \over 2k}e^{-k|z|}e^{ik(x-ct)}\, .
\ee
In Figure 3 the velocity field associated with $\psi$ is plotted. The evanescent structure of the vector field away from the interface shows that both the divergence and the vorticity fields are zero everywhere ($\partial u/ \partial x = -\partial w/\partial z$ and $\partial w/\partial x = \partial u/\partial z$) except at the interface, where the vorticity goes to infinity in magnitude since the jump in the sign of $u$ across the interface generates the vortex sheet. Since $u(0^-)=-u(0^+)$, we have $u_0=0$. { This implies that the non-Boussinesq baroclinic torque term in Equation \ref{eq:2.10} vanishes in the DW limit}. Furthermore, from Equation \ref{eq:3.1} we obtain, ${\hat w}_0 = -i{{\hat q}_0/ 2}$, thus Equations \ref{eq:2.15a}--\ref{eq:2.15b} gives the familiar DW phase speed relation:
\be
%\[c =  {{\hat q}_0\over 2k{\hat \eta}_0}   = {2g{\hat \eta}_0 \over {\hat q}_0} = \pm \sqrt{g\over k} \]_{DW}\, ,
c =  {{\hat q}_0\over 2k{\hat \eta}_0}   = {2g{\hat \eta}_0 \over {\hat q}_0} = \pm \sqrt{g\over k}\, ,
\ee
corresponding to:
\be
%\[\({{\hat q}_0\over {\hat \eta}_0}\)^{\pm} = \pm 2\sqrt{gk} \]_{DW}\, . 
\({{\hat q}_0\over {\hat \eta}_0}\)^{\pm} = \pm 2\sqrt{gk} \, .
\ee
Hence, when the interface displacement and vorticity are in (anti) phase the wave propagates to the (left) right (see Figure \ref{fig:3}). Such behavior is common with other types of interfacial vorticity waves such as capillary \citep{biancofiore2017}, Rossby \citep{hoskins1985use} and Alfven \citep{heifetz2015interacting} waves.

{ We note here that although we are considering a ``non-Boussinesq'' density jump (air-water interface), the non-Boussinesq effect is \emph{completely absent} for the deep-water limit. In other words, this implies that the Boussinesq effects are not confined to ``small density variations'', as is conventionally understood.} A similar conclusion has been obtained by  \cite{guha2018inertial}  using detailed scaling arguments.    

\subsection{Finite depth}

\begin{figure}
\centering
\includegraphics[width=1.0\linewidth]{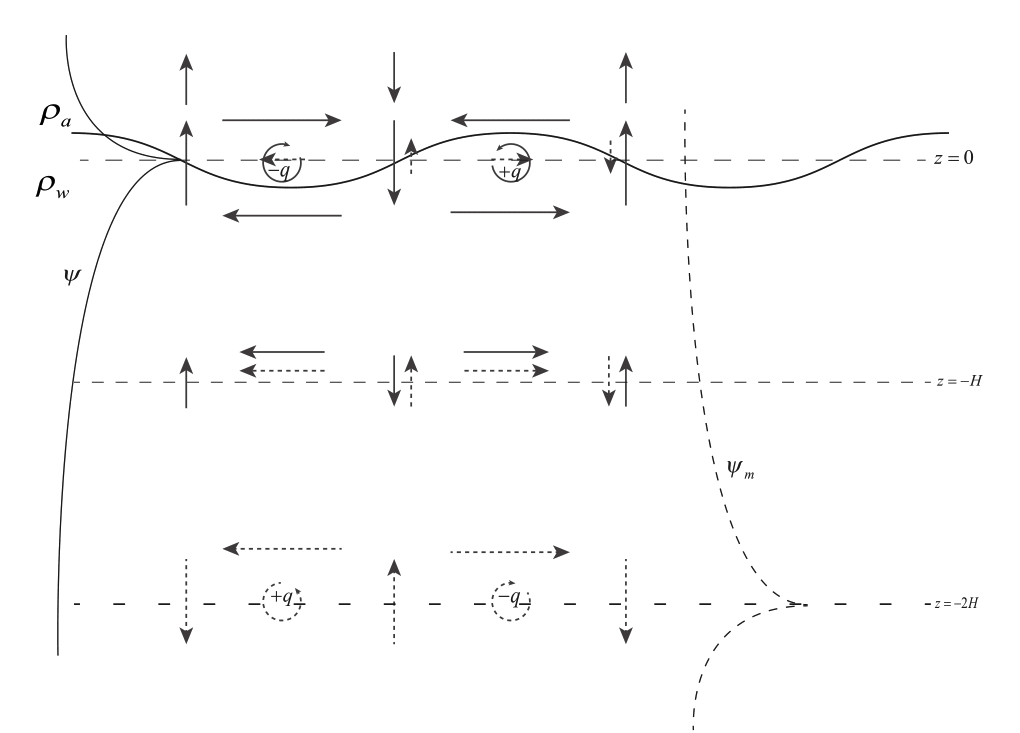}
\caption{Same as in Figure \ref{fig:3} but for the finite depth case. Here, the velocity field
 results  from Equation \ref{eq:3.4}, which is the superposition of the deep-water streamfunction (indicated  in Figure \ref{fig:3} by the solid evanescent profile on the left) and its
anti-phase mirror image streamfunction $\psi_m$, whose evanescent structure away from
$z = -2H$ is shown on the right with dashed line. The superposition of the induced vertical velocity from the air-water interface
and its mirror image completely cancel each other at $z =-H$, thereby satisfying impenetrability. Moreover,
the vertical velocity induced by the mirror image on the interface at $z = 0$ is in
anti-phase with the local self-induced vertical velocity. Therefore, according to
Equation \ref{eq:2.15a} and the illustration in Figure \ref{fig:2}, this slows the interface propagation. Furthermore, the horizontal velocity at the interface is  non-zero (unlike the deep-water case) since it has a
contribution from the mirror image, which is in phase with the vorticity anomaly
at the interface. This stands in agreement with Equation \ref{eq:2.15b} as it decreases the
phase speed magnitude. The latter effect is due to the non-Boussinesq baroclinic
torque (c.f. Equations \ref{eq:2.09}--\ref{eq:2.10} and Figure \ref{fig:1}b).
}
\label{fig:4}
\end{figure}

In the finite depth case, the effect of the impenetrability of the solid boundary, located at a finite distance ($z=-H$), can be represented equivalently by replacing that boundary by an anti-phased undulating vortex sheet (mirror image of the one at $z=0$), located at $z=-2H$. This yields
\be
\label{eq:3.4}
\psi = -{{\hat q}_0 \over 2k}\left(e^{-k|z|}-e^{-k|z+2H|}\right)e^{ik(x-ct)},\, 
\ee
which indeed vanishes at $z=-H$, satisfying impenetrability. Furthermore, this anti-phased mirror vortex sheet reduces the vertical velocity at the interface at $z=0$ as it induces an evanescent vertical velocity of opposite sign (see Figure \ref{fig:4}):
\be
{\hat w}_0 = -i{{\hat q}_0\over 2}\left(1- e^{-2kH}\right)\, .
\ee
We readily observe that when $kH \rightarrow \infty$, we recover  ${\hat w}_0 = -i{{\hat q}_0/ 2}$, i.e., the vertical velocity in the deep-water limit. Due to the presence of a solid boundary at a finite depth, the phase speed should decrease according to Equation \ref{eq:2.15a}.
% Therefore, according to Equation (2.15a), the phase speed should decrease due to the presence of a solid boundary at finite depth. 
This reduction in the phase speed magnitude is obtained from Equation \ref{eq:2.15b} as well. Indeed, as illustrated in Figure 4, the anti-phased mirror image vortex sheet at $z=-2H$ induces a horizontal velocity $u$ which is in phase with the vortex sheet at $z=0$. In fact, the sole contribution to $u_0$ is from the anti-phased mirror image vortex sheet. From Equation \ref{eq:3.4}, $\hat{u}_0$ is found to be
\be
\label{eq:3.6}
\hat{u}_0 = -\(\der{\psi}{z}\)_{z=0} = {{\hat q}_0 \over 2}e^{-2kH}\,.
\ee
{
Substituting Equation \ref{eq:3.6} back in Equation \ref{eq:2.10} we obtain \be
\label{eq:vor_prod}
%\(1 + e^{-2kH}\)\der{q}{t} = -N^2\der{\eta}{x},
\der{q}{t} = -\frac{N^2}{(1 + e^{-2kH})}\der{\eta}{x}.
\ee
%where the non-Boussinesq torque results in the second term of the bracket in the LHS. 
Hence the interface displacement slope, $-\partial{\eta}/\partial {x}$, is able to produce less  vorticity ($\partial q/\partial t$); the non-Boussinesq term $e^{-2kH}$
 appearing in the denominator of the RHS is always greater than $1$.
%Hence the vorticity production by the interface displacement slope, $-\partial{\eta}/\partial {x}$, becomes smaller when the non-Boussinesq term is taken into account.
}
Stated differently, the two terms in the RHS of Equation \ref{eq:2.09} work  against each other (while the first Boussinesq term $T_1$ dominates the second non-Boussinesq term $T_2$), since now the interface displacement correlates with the pressure perturbation there (c.f. Figure  \ref{fig:1} and Figure  \ref{fig:4}). 
Substituting $({\hat u}_0, {\hat w}_0)$ in Equations \ref{eq:2.15a}--\ref{eq:2.15b} gives
\be
\label{eq:3.7}
c =  {{\hat q}_0\over 2k{\hat \eta}_0}\(1-  e^{-2kH}\)  = {2g{\hat \eta}_0 \over {\hat q}_0 \(1 + e^{-2kH}\)} = \pm \sqrt{{g\over k}\tanh{(kH)}}\, , 
\ee
corresponding to:
\be
\label{eq:3.8}
\({{\hat q}_0\over {\hat \eta}_0}\)^{\pm} = \pm 2\sqrt{gk\over 1-  e^{-4kH}}\, . 
\ee
 Equation \ref{eq:3.7} is the familiar dispersion relation of intermediate depth surface gravity waves.

\subsection{Shallow-water limit}
\label{subsec3_2}

\begin{figure}
\centering
\includegraphics[width=1.0\linewidth]{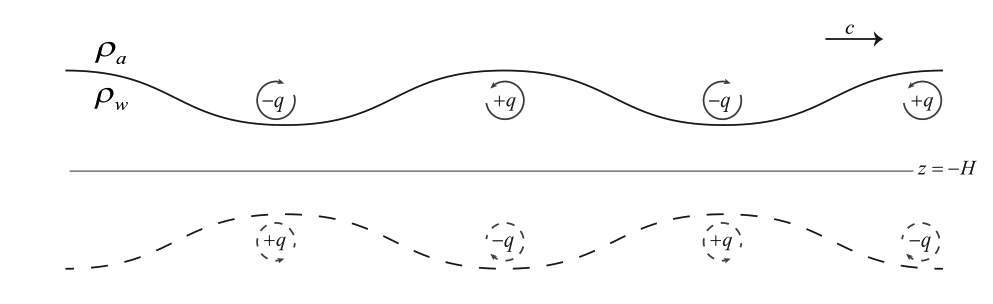}
\caption{The wave structure for the shallow-water limit can be viewed as a
series of dipole-like bulges due to the vorticity interface and its anti-phased mirror
image. In this illustration the interface anomaly is highly exaggerated in order
to show the effect. When the wave propagates to the right (as shown here),
the positive vorticity anomaly is above the negative one within the bulges. For
leftward propagation (not shown here) the structure is upside down.
}
\label{fig:5}
\end{figure}

The shallow-water (hereafter SW) limit of $kH \ll 1$ is also interesting to analyze from this perspective. The  vertical structure of the streamfunction becomes independent of the wavenumber when the vortex sheet and its anti-phased mirror image become very close to each other, thereby generating a series of dipole-like structures (Figure \ref{fig:5}). This reflects in the phase speed, which now becomes independent of the wavenumber. As can be seen from Equation \ref{eq:3.4}, under the shallow-water limit
\be
%\[\psi = {{\hat q}_0 \over 2}\left(|z|-|z+2H|\right)e^{ik(x-ct)} \]_{SW}\, ,
\psi = {{\hat q}_0 \over 2}\left(|z|-|z+2H|\right)e^{ik(x-ct)}\, ,
\ee
yielding $({\hat u}_0,{\hat w}_0) = ({1\over 2}, -ikH){\hat q}_0$. Substitution of this in Equations \ref{eq:2.15a}--\ref{eq:2.15b}  gives
\be
\label{eq:3.10}
%\[c =  H{{\hat q}_0\over {\hat \eta}_0}   = g{{\hat \eta}_0 \over {\hat q}_0} = \pm \sqrt{gH} \]_{SW}\, , 
c =  H{{\hat q}_0\over {\hat \eta}_0}   = g{{\hat \eta}_0 \over {\hat q}_0} = \pm \sqrt{gH} \, , 
\ee
corresponding to:
\be
%\[\({{\hat q}_0\over {\hat \eta}_0}\)^{\pm} = \pm \sqrt{g\over H} \]_{SW}\, . 
\({{\hat q}_0\over {\hat \eta}_0}\)^{\pm} = \pm \sqrt{g\over H}\, . 
\ee
Again, one can easily recognize that Equation \ref{eq:3.10} is the familiar dispersion relation for shallow-water waves.
{It is straightforward to verify that in the SW limit the magnitude of the opposing non-Boussinesq torque is equal to half of the Boussinesq one.}

%We wish to provide some intuition to help understanding why the gravity waves become non-dispersive in the shallow-water limit.
%Since $u$ is not a function of $z$ in this limit, vertical integration of the continuity equation assures that the amplitude of $w$ at the interface is proportional to the wave number and to the vorticity interface anomaly that induces the motion, i.e.,  $|{\hat w}_0| \propto k|{\hat q}_0|$. Thus, for the same value of vorticity amplitude, $|{\hat w}_0|$ is inversely proportional to the wavelength. On the other hand $|{\hat w}_0| = k|c {\hat \eta}_0|$, which together with the previous relation imply that $|c {\hat \eta}_0| \propto |{\hat q}_0|$. Furthermore, since in the shallow water limit we obtain as well that the ratio between the values of the interface vorticity and displacement, $|{\hat q}_0/{\hat \eta}_0|$, is not a function of the wavelength, $c$ becomes non-dispresive. Stated differently, the inversely proportional evanescence of the vertical velocity with the wavelength shifts the interface displacement, either to the right or to the left, at a speed that is independent of the wavelength itself

\section{Gravity wave energy from the interfacial fields}
\label{sec4}
\setcounter{equation}{0}

Defining $K \equiv {{\overline \rho}}\left(u^2 + w^2\right)/2$ as the kinetic and $P \equiv{\overline \rho} (N\eta)^2/2$ as the potential energies, it is straightforward to deduce from Equations \ref{eq:3d}--\ref{eq:3b} that
%From equation set (3) it is straightforward to deduce that:
\be
\label{eq:4.1}
\der{}{t}\<K\> = -\<w\rho g\> = - \der{}{t}\<P\>\, ,
\ee
where the domain integration operator is defined as
\be
\<\dots\> \equiv {1\over \lambda}\int_x^{x+\lambda}\int_{z=-H}^{\infty} (\dots) dz dx\, ,
\ee
% and the kinetic and potential energy respectively are:
% \renewcommand{\theequation}{A.\arabic{equation}a,b}
% \be
% K = {1 \over 2}{{\overline \rho}}\left(u^2 + w^2\right)\, ; \hspace{0.5cm}
% P = {1 \over 2} {\overline \rho} (N\eta)^2\, .
% \ee
where $\lambda=2\pi/k$ is the wavelength. 
Equation \ref{eq:4.1} indicates that the total wave energy $\<E\> = \<K+P\>$ is conserved.
After integration by parts it can be shown that
\be
\<K\> = -{1\over 2}\<\psi\({\overline \rho} q - \deriv{\overline \rho}{z}u\)\>\, .
\ee
Substituting Equations \ref{eq:2.12}, \ref{eq:3.4} and \ref{eq:3.6} we obtain
\be
\<K\> ={\rho_w \over 16 k} \(1-  e^{-4kH}\){\hat q}^2_0\, ,
\ee
and using Equation \ref{eq:3.8} we obtain the familiar result
\be
\<K\> = {1\over 4} \rho_w g {\hat \eta}^2_0\,.
\ee
Finally, substituting $N^2 = 2g\delta(z)$ in $P \equiv{\overline \rho} (N\eta)^2/2$ we recover the well known equi-partition energy relation: $\<K\> = \<P\>$, as demanded by the Virial theorem \citep{buhler2014waves}.

\section{Discussion}
\label{sec5}

%This short analysis emerged from a feeling of unsatisfactory which arose by us when trying to explain undergraduate students (and ourselves) the essence of the physical propagation mechanism of surface gravity waves. The standard potential flow derivation is straightforward mathematically, however it provides an impression that the flow is irrotational everywhere, and furthermore it is difficult to assess whether or not the underlying dynamics is a Boussinesq one. 

 {In geophysical fluid dynamics, gravity waves are often associated with the fast divergent component of the flow, whereas the vortical component is associated with the slower quasi-balanced Rossby wave-like dynamics. For the latter, the vertical component of  vorticity (or potential vorticity) equation plays the key role in understanding its evolution and vorticity inversion is used to obtain the far field velocity field induced by the vorticity field. The fact that interfacial gravity waves, generated by density discontinuities, can be treated as undulated vortex sheets has been acknowledged a while ago  (e.g. \cite{baker1982generalized}), however to the best of our knowledge, its spanwise (i.e. the meridional direction for zonal wave propagation) vorticity equation has not been implemented to understand the propagation mechanism of gravity waves.}  

Analysis of the spanwise vorticity equation reveals that the interfacial vorticity is translated by the baroclinic torque acting on the interface when the dynamics deviates from the rest state of hydrostatic balance. {The baroclinic torque has two components - Boussinesq and non-Boussinesq, which tend to act against each other, where the Boussinesq always overwhelms the non-Boussinesq.} The non-Boussinesq component depends on the perturbation horizontal pressure gradient force  across the interface, which in turn accelerates the zonal perturbation velocity there. Using vorticity inversion and mirror imaging technique for the bottom boundary, we see that the zonal velocity at the interface results solely from the vorticity induced by the mirror imaging. 
{Therefore, in the deep water limit, this contribution vanishes. Hence, the non-Boussinesq dynamics vanishes despite of the large density contrast at the water-air interface. When the water depth decreases, the magnitude ratio of the non-Boussinesq and Boussinesq baroclinic torques increases, reaching the value of $1/2$ in the shallow water limit.} The same dispersion relation is obviously obtained if  the conventional dynamic condition of the time-dependent Bernoulli equation is use instead, however it is not clear from the latter how to disentangle the non-Boussinesq dynamics from the Boussinesq one.

{Even for infinite depth, the non-Boussinesq torque should still be taken into account when multiple density interfaces exist \citep{heifetz2015stratified,guha2018inertial}. This is because the vortex sheet gravity waves induce far field horizontal fields on each other and thereby activate the non-Boussinesq dynamics. As opposed to the mirror images, which are always in anti-phase with the vortex sheet interface,  the gravity waves moves relative to each other when background shear is present. Thus a non-Boussinesq stratified shear flow action at a distance between counter-propagating interfacial gravity waves can lead to a phase locking resonant instability in a similar (though more complex) manner of the barotropic and baroclinic instability between Rossby waves, described in \cite{hoskins1985use}}.

{The undulating vortex sheet understanding of surface gravity waves shares some similarities with the Bretherton  delta function potential vorticity representation of the solid boundary condition in baroclinic quasi-geostrophic  setups \citep{bret1966}. Specifically, here we showed that the total, domain integrated  energy of the gravity waves can be obtained solely from evaluating the dynamical fields at the air-water interface, similar to the evaluation of the total energy of the discrete spectrum in the Eady model solely from the boundary fields \citep{eady1949long,heifetz2004counter}.

{\bf {Acknowledgments}} We are very grateful to Jeff Carpenter from HZG and Raunak Raj from IIT Kanpur for insightful comments. A.G. gratefully acknowledges Alexander von Humboldt research fellowship for funding support. The authors also thank the anonymous reviewers for helpful comments and suggestions. \\

\bibliographystyle{wileyqj} 

 \bibliography{paper1}

\end{document}